\documentclass[
 aip,
 amsmath,amssymb,
 reprint,%
]{revtex4-1}

\usepackage{graphicx}
\usepackage{dcolumn}
\usepackage{bm}

\usepackage[utf8]{inputenc}
\usepackage[T1]{fontenc}
\usepackage{mathptmx}

\begin{document}

\preprint{AIP/123-QED}

\title{Suppression of Brillouin oscillation in transparent free-standing diamond thin films in picosecond ultrasound\\}

\author{H. K. Weng}
\affiliation{Graduate School of Engineering, Osaka University, 2-1 Yamadaoka, Suita, Osaka 565-0871, Japan}
\author{A. Nagakubo}
\affiliation{Graduate School of Engineering, Osaka University, 2-1 Yamadaoka, Suita, Osaka 565-0871, Japan}
\author{H. Watanabe}
\affiliation{National Institute of Advanced Industrial Science and Technology, Tsukuba, Ibaraki 305-8568, Japan}
\author{H. Ogi}
\email[]{ogi@prec.eng.osaka-u.ac.jp\\
 Accepted for publication in \textit{Applied Physics Letters}}

\affiliation{Graduate School of Engineering, Osaka University, 2-1 Yamadaoka, Suita, Osaka 565-0871, Japan}

\date{\today}

\begin{abstract}
Brillouin oscillation appears in picosecond ultrasonics for a transparent specimen because of backward light scattering by moving strain pulse.  Its amplitude is comparable with those of other responses, such as pulse-echo signals and through-thickness resonance, obscuring these non-Brillouin-oscillation responses. We here find that Brillouin oscillation can be suppressed in a transparent free-standing film by coating both sides with metallic thin film of appropriate thickness and that this peculiar behavior is caused by strain pulses generated on both sides with a slight phase difference. This phenomenon allowed us to fabricate a Brillouin-oscillation-free diamond free-standing film, which showed high capability for sensor applications.
\end{abstract}

\maketitle
Picosecond ultrasound \cite{Thomsen,Grahn,Stoner,Matsuda} is a key methodology for studying ultrahigh-frequency coherent-phonon behavior in thin films. It uses pump and probe light pulses for generation and detection of coherent phonons, respectively.  The longitudinal-wave ultrasound is generated by transient thermal stress near the surface, and it is detected through the reflectivity change in the probe light mainly caused by the piezo-optic effect. This technique is highly promising for evaluating ultrasound attenuation in thin films,\cite{Zhu,Devos1,Emery,Devos2} which is an important parameter for designing ultrahigh-frequency ultrasonic-resonator devices such as bulk-acoustic-wave resonators.\cite{Ueda,Thalmayr,Hara} The most reliable attenuation measurement can be made on a free-standing thin film because of no energy leakage to the substrate.  The free-standing thin film is also suitable for sensor applications, including mass-sensitive biosensors. Quartz crystal microbalance (QCM) biosensor is a representative mass-sensitive biosensor, \cite{Muramatsu} which detects targets using the mass-loading effect on the resonator through the change in the resonance frequency. Because the mass sensitivity is inversely proportional to the square of the thickness of the resonator, \cite{Sauerbrey} a thin-film resonator is expected to improve the sensitivity. \cite{OgiApex, OgiBB}  A free-standing thin-film resonator is therefore promising to achieve an ultra-sensitive biosensor relying on its high quality factor.  

For an opaque free-standing thin film, the ultrasonic pulse generated near the surface causes the pulse-echo signals or the through-thickness resonance. However, the response from a transparent free-standing thin film becomes more complicated because of the Brillouin oscillation, \cite{Devos1,Matsuda,Ogi2008PRB} which is caused by the backward scattering of the probe light due to the moving strain pulse. The Brillouin-oscillation amplitude sometimes exceeds those of non-Brillouin-oscillation responses, deteriorating their measurement accuracy. Therefore, disappearing the Brillouin oscillation is an unsolved problem in picosecond ultrasound. Here, we show that the Brillouin oscillation can be suppressed in a transparent free-standing thin film.  We furthermore applied such a free-standing thin-film resonator to a biosensor application.

\begin{figure}[tb]
\begin{center}
\includegraphics[width=80mm]{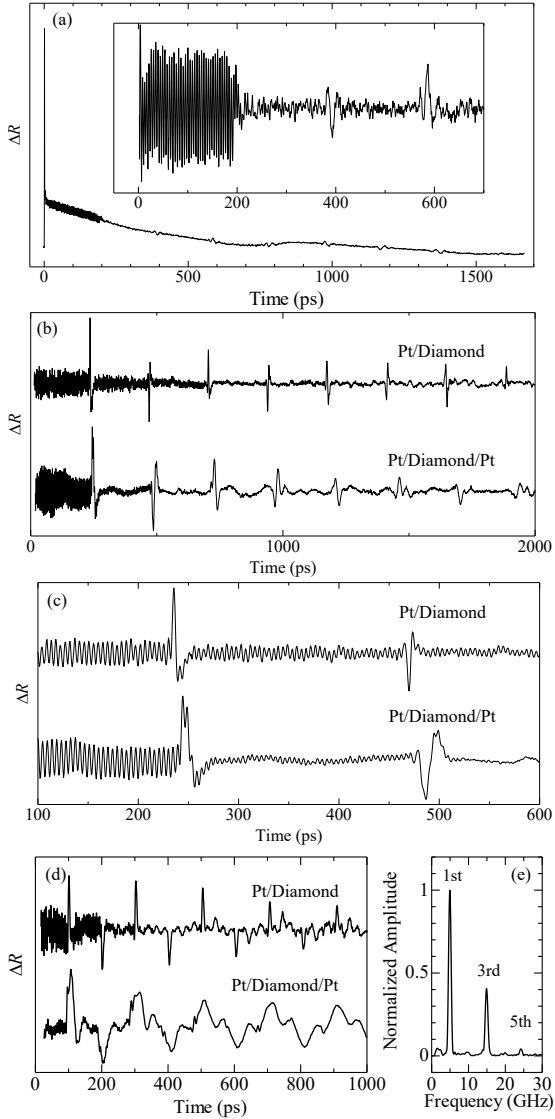}
\caption{(a) As measured reflectivity change of the probe light for a free-standing diamond film (3.5 $\mu$m thick) with 10-nm Pt thin film on both sides. The inset shows the background subtracted reflectivity change near the zero point. (b) Measurements for 4 $\mu$m thick free-standing diamond thin film with Pt thin film only on the single side (upper) and on both sides (lower). (c) Enlarged reflectivity changes near the first and second echoes in (b). (d) Measurement for 1.7 $\mu$m thick diamond thin film with Pt film on the single side (upper) and both sides (lower). (e) The FFT spectrum of the lower data of (d).}
\end{center}
\end{figure}

We used a typical picosecond ultrasound method in this study. A Ti-sapphire pulse laser with 140 fs duration and 80 MHz repetition rate was used. The wavelengths of pump and probe lights were 800 and 400 nm, respectively. The details of our optics are shown elsewhere. \cite{Nagakubo2019,Weng2020,Weng2021}

We prepared a 350-nm graphite thin film and diamond free-standing thin films. The edge of each specimen was supported by a 0.5-mm silicon plate, remaining the free-standing portion. The graphite free-standing film was synthesized by heating a polyimide film at $\sim$3000 K to grow grains up to 20 $\mu$m. \cite{Kusakabe} Because our microscopic measurement system allows measurements in local regions of $\sim$1 $\mu$m, \cite{NagakuboAPE} it is possible to measure the phonon properties inside a single grain. \cite{Kusakabe}  

Concerning the free-standing diamond specimens, we first synthesized  the homoepitaxial [100] diamond thin film on a 50-$\mu$m thick [100] diamond substrate by the microwave plasma-assisted chemical-vapor-deposition (CVD) method. \cite{Costello,Watanabe,Weng2021} The substrate was then removed by gas etching method to obtain the free-standing monocrystal-diamond thin films. Their thicknesses are between 1.7 and 4 $\mu$m. A 10-nm Pt thin film was then deposited by the magnetron sputtering method on each specimen for the ultrasound generation. 

Figure S1 shows the measurement on the graphite specimen. The pulse-echo signal appears every 155 ps, and considering the sound velocity of about 4.5 nm/ps, \cite{Kusakabe} it appears every round trip.  The echoes are in phase because of reflections at free boundaries.

On the other hand, the signals from the free-standing diamond thin films are significantly different from that from the graphite (opaque) free standing film as shown in  Fig.1. First, the phase of the pulse echo changes every time despite the reflection at the free boundary. Second, the time interval between nearby echoes corresponds to the single trip, not the round trip. Similar observation \cite{Grossmann} has been reported for a $\sim$300-nm free-standing Si film with a 820-nm light (Si thin film is a nearly transparent material at this wavelength). Third, the through-thickness resonance vibrations are observed when the Pt film is deposited on both sides for a thinner film, as shown in the lower data in Fig. 1(d), and its fast-Fourier-transform (FFT) spectrum (Fig. 1(e)) indicates that only odd-numbered modes are excited. Similar observations are reported for free-standing Si films, \cite{Grossmann,Bruchhausen,Hudert} although the excitation mechanism for the odd resonance mode is different as described later. 

Here, we find an unusual phenomenon: The Brillouin oscillation in diamond is occasionally highly weakened (almost disappearing) after the first echo when the Pt thin film is deposited on both sides, as shown in Figs. 1(a) and lower data in Figs. 1(b)-(d).  Possible mechanisms for the acoustic-wave generation in the free-standing diamond film with the Pt thin film will originate from the thermal stress, $-3B\beta\Delta T$, in the Pt layer, and the electronic stress due to the deformation potential, \cite{Grahn, Wright} $-BN(\partial E_g/\partial P)$, in the diamond layer.  Here, $B$, $N$, $E_g$, $P$, $\beta$, and $\Delta T$ denote the bulk modulus, density of the electron-hole pair generated by the pump light, band-gap energy, pressure, thermal expansion coefficient, and the temperature increment, respectively.  (The stress caused by the thermoelastic effect in the diamond layer due to the coupling between electrons and phonons will be negligible because of the pump-light energy lower than the band-gap energy.) \cite{Chuan} Using literature values for these parameters (for example, $\partial E_g/\partial P$=0.07 eV/Mbar), \cite{Camp} it is found that the thermal stress in the Pt layer is more than two orders of magnitude larger than the electronic stress in the diamond layer, allowing us to consider the former as the principal wave-generation source. 

In order to explain the suppression of the Brillouin oscillation in the diamond free-standing films, we calculate the reflectivity change numerically based on theories developed previously. \cite{Zhu,Ogi2008PRB} The reflectivity change is obtained by
\begin{equation}
\Delta R(t)=\left | r_0(t)+\Delta r(t) \right |^2-\left | r_0(0) \right |^2,
\label{DR}
\end{equation}
where $r_0(t)$ and $\Delta r(t)$ denote the reflection coefficient from the thin film and its variation due to the photoelastic effect, respectively, and they are given by

\begin{eqnarray}
r_0(t)=\frac{r_{01}+r_{10}e^{-2ikd(t)}}{1+r_{01}r_{10}e^{-2ikd(t)}},
\label{GL}
\end{eqnarray}
\begin{equation}
\Delta r(t)=\int_{0}^{d(t)}g(z,t)\eta(z,t)dz.
\end{equation}
Here, $r_{01}$ and $r_{10}$ are the reflection coefficients from the vacuum to the thin film and from the thin film to the vacuum, respectively. $k$ denotes the wavenumber of the probe light in the diamond film.  $d(t)$ denotes the total thickness of the thin film, which is time-dependent because it is varied by the strain pulse, $\eta(z,t)$. $g(z,t)$ indicates the sensitivity function, which takes the following form for a free-standing film: \cite{Ogi2008PRB}

\begin{equation}
g(z,t)=-\frac{i}{2}\frac{k_0^2}{k} \frac{t_{01}t_{10}e^{-2ikz}(1+r_{10}e^{-2ik(d(t)-z)})^2}{1+r_{01}r_{10}e^{-2ikd(t)}}p,
\end{equation}
where $t_{01}$ and $t_{10}$ are the transmission coefficients from the vacuum to the thin film and the reverse, respectively.  $p$ denotes the piezo-optic coefficient in the thin film. Note that $r_0(t)$ is time-dependent because the total thickness is varied by the strain pulse, indicating that the reflectivity change will be detected even without the photoelastic effect ($p$=0) because of the change in the optical path.   

The strain pulse generated by the pump light pulse is estimated by \cite{Zhu}
\begin{equation}
\eta(z,t)=-\eta_0\textrm{exp}(-\left | z-vt \right |/\zeta)\textrm{sgn}(z-vt),
\end{equation}
where $\eta_0$ and $\zeta$ are the strain amplitude and the strain pulse width, respectively. We simply used the Pt thickness for $\zeta$ and assumed the generation of the strain pulse directly on the diamond thin film, and performed the numerical simulation without propagation in the Pt layer. Because the acoustic impedance of Pt is close to that of diamond (the acoustic transmittance of ~80\%), the ringing of the Pt film is insignificant if any. (Such a ringing of Pt film is not observed clearly in experiments.)  We calculated the reflectivity change $\Delta R(t)$ using the following procedure.  First, the strain pulse with frequency-dependent (wavenumber-dependent) attenuation at a propagation distance $\xi(=vt)$ was obtained by reducing the amplitude of the Fourier component using the corresponding attenuation value determined by exp(-$\alpha \xi$) and reconstructing the strain pulse using the inverse Fourier transform. $\alpha$ indicates the attenuation coefficient, and we assume that it is proportional to the square of the frequency because of the anharmonicity process \cite{Woodruff}
\begin{equation}
\alpha(\omega)=\beta\omega^2,
\end{equation}
where $\beta$ denotes the coefficient. Second, the total thickness $d(t)$ was calculated by integrating the strain.  Third, the reflectivity change was calculated by equations (1)-(3).

Figure 2 (a) shows a simulation result for a 4-$\mu$m thick free-standing diamond thin film, when the pump light pulse is absorbed only on the top surface.  We can see strong Brillouin oscillation because of the transparency of the material, as well as the pulse-echo signal that appears every single-trip propagation.  Note that the phase reversal in the pulse-echo signal is reproduced.  One may attribute this to the effect of the total thickness change on the reflection coefficient $r_0(t)$, because the phase of the strain pulse is reversed at the free end, inducing the opposite thickness change at reflections between bottom and top surfaces.  However, we find that this effect is negligible compared by the photoelastic contribution (Eq. (3)) because of extremely large elastic constant of diamond.  Equation (3) indicates that the reflectivity change related to the photoelastic effect is governed by the integral of the strain-pulse-related quantity over the thickness, which implicitly includes the total thickness change.  During propagation inside the film, the film thickness will be nearly the same as that without the strain pulse due to the bipolar character of the strain pulse.  However, when the strain pulse reflects at the free end, the total thickness changes transiently, which can be detected through $\Delta r(t)$.  

\begin{figure}[tb]
\begin{center}
\includegraphics[width=75mm]{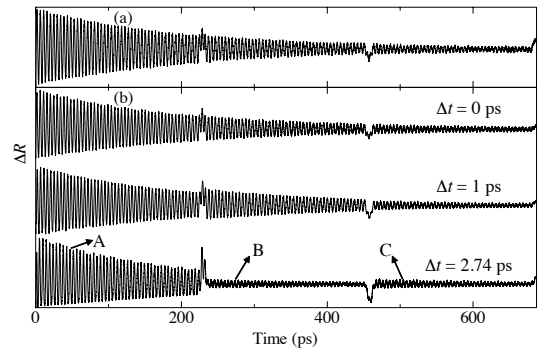}
\caption{Numerical calculation for the probe-light reflectivity change in 4-$\mu$m thick free-standing diamond thin film (a) when the strain pulse is generated from only the top surface and (b) when the strain pulses are generated from both sides with a time delay $\Delta t$. Used parameters are as follows: $v$=17.5 nm/ps, \cite{Weng2021} $n$=2.46, \cite{Palik} $\zeta$=10 nm, $\lambda_0$=400 nm, and $\beta$=0.00015 nm$^{-1}$THz$^{-2}$. The Brillouin-oscillatioin frequency $f$ is given by $f=2nv/\lambda_0$, which equals 215 GHz.  This value is in good agreement with the experiment (214.3 GHz) in Figs. 1(a) and (b). (Waveforms for the two strains at three time points A, B, and C are shown in Figure S2.) Supplementary Movie 1 ($\Delta t$=1 ps) and Movie 2 ($\Delta t$=2.74 ps) show the relationship between the reflectivity change and the strain pulses in more detail.}
\end{center}
\end{figure}

Concerning the sudden reduction of the Brillouin-oscillation amplitude after the first echo observed for diamond free-standing thin films, we first considered the effect of surface roughness. Because the Brillouin-oscillation frequency in diamond is very high (215 GHz), a scattering loss would be caused at the reflection because the amplitude loss due to surface roughness is proportional to the square of the frequency, \cite{Nagy} so that the high-frequency component would be diminished at each reflection. However, the rms surface roughness on our diamond specimens is about 1 nm both before and after the Pt-film deposition, which was confirmed by an AFM measurement, and this value is much smaller than the ultrasound wavelength (81 nm), so we do not expect significant amplitude loss at the reflection. Moreover, the reduction of the Brillouin-oscillation amplitude was observed when the Pt film was additionally deposited on the back side (lower data in Figs. 1(b) and (c)), and this behavior will not be explained by the surface roughness.  Ultrasonic absorption in the Pt layer might be another possible factor.  However, this is insignificant because of two reasons. First, the absorption loss occurs when the material is strained, but the strain in the Pt film is expected to be very small because of the free surface condition. Second, if it were a principal cause, suppression of Brillouin oscillation would have been observed even in the diamond with single Pt layer. 

We attribute the additional generation of the strain pulse at the back interface between diamond and Pt to the unusual response. As the Pt film is very thin, the pump light pulse is able to pass through it and reach the bottom surface of the diamond film, exciting the strain pulse at the interface as well. The double peaks observed at the first echo (at 250 ps of the lower data in Fig. 1(c)) indicate that another strain pulse propagates in the specimen with a slight phase difference. Excitation of the strain pulse at the back side also consistently explains the through-thickness resonance observed in the Pt/Diamond/Pt free-standing specimen (lower data in Fig. 1(d)), which was not observed in the single-side excitation (upper data in Fig. 1(d)), since the thermal stress on both sides will enhance the odd through-thickness resonances because of the deformation symmetry as shown in Fig. 1(e).  

We thus assume that the extra strain pulse is generated at the back interface. We performed the numerical simulation of the reflectivity change caused by propagations of the two strain pulses. Because of the difference in the generation point, there will be a phase difference between the two strain pulses: The pump light pulse is expected to generate the strain pulses near the outer surface of the top Pt film and near the inner surface of the bottom Pt film, simultaneously. The light-pulse power for the latter should be lower than that for the former. However, the inner surface of the bottom Pt film is not a free end unlike the outer surface of the upper Pt film, and the thermal stress there will be enhanced because of the restriction of deformation.  We then simply assumed the same strain amplitude for these two strain pulses based on these two opposite contributions. The time difference between them ($\Delta t$) may correspond to the propagation time of the strain generated at the upper surface of the Pt layer, which is estimated to be 2.4 ps with the wave speed of 4.2 nm/ps in Pt. Although it is not straightforward to take the fast diffusion of the hot electrons\cite{Tas,Block} into account, the thermal stress generated during the electron thermalization will be greater near the outer surface in the upper Pt layer and near the interface in the bottom Pt layer. Therefore, the effective starting points of the strain pulses will be different between the upper and bottom Pt layers. 

Figure 2(b) presents the simulation results. Note that if there is no time difference between the two strain pulses, the result is nearly identical to the case that the single pulse is generated only from the front surface (Fig. 2(a)), and we fail to reproduce the Brillouin-oscillation disappearance.  With a time difference, however, the amplitude of Brillouin oscillation after the first pulse echo can be modified. When $\Delta t$ = 2.74 ps, it is significantly decreased, which is similar to the experiments. This reduction occurs because of the interference between diffracted probe lights by the strain pulses propagating in the opposite directions.  The double peaks in the first pulse are also reproduced. The amplitude recovers after the second pulse, but the material attenuation remains the low amplitude. (Supplementary Movies 1 and 2 show the relationship between the reflectivity change and the strain propagation in more detail when $\Delta t$ = 1 and  2.74 ps, respectively.) Importantly, this time difference is close to the estimation of 2.4 ps, supporting our view.  The discrepancy (14\%) may be attributed to the ambiguity of the film thickness, the elastic constants of Pt thin film \cite{Nakamura,Ogi2007}, and the diffusion behavior of the hot electrons. 

We investigate the effect of the time difference and the film thickness on the reduction of the Brillouin-oscillation amplitude.  Figure S3 presents the simulation results for the ratio of the Brillouin-oscillation amplitudes before and after the first-echo ($A_{\textrm{after}}/A_{\textrm{before}}$) without attenuation.  It takes a maximum at $\Delta t=\sim$2.1 ps and a minimum at $\Delta t=\sim$2.7 ps (Fig. S3(a)).  Also, the difference in film thickness where the amplitude ratio takes maximum and minimum is only about 40 nm (Fig. S3(b)).  Therefore, the amplitude ratio is very sensitive to the thicknesses of the Pt layer and the diamond film.  These results consistently explain the experimental fact that the suppression of the Brillouin oscillation does not always happen (Fig. S4) because a very small thickness difference can affect significantly the  Brillouin oscillation after the first pulse echo.

\begin{figure}[tb]
\begin{center}
\includegraphics[width=85mm]{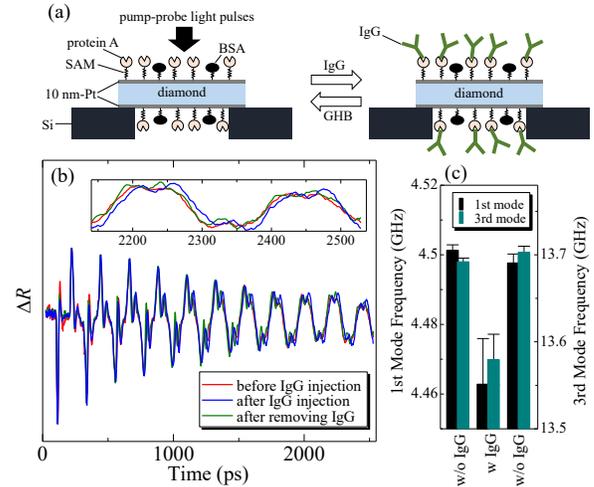}
\caption{(a) Illustrations of modification of surfaces of the diamond free-standing thin film resonator.   Bovine serum albumin (BSA) was used for blocking the non-specific binding of the IgG molecules.  (b) Reflectivity changes measured at the three stages: Before and after exposed to the IgG solution, and after the rinsing with a glycine-HCl buffer (GHB).  The inset shows a magnified view of the reflectivity change.  (c) Changes in the resonant frequencies of 1st and 3rd modes at the three stages.}
\end{center}
\end{figure}

We have thus developed a free-standing diamond film with the Pt thin film on both sides, where Brillouin oscillation is sufficiently suppressed, and applied it to a biosensor to demonstrate the importance of this phenomenon.  We immobilized protein A molecules on both surfaces using a self-assembled monolayer (SAM) for capturing immunoglobulin G (IgG) relying on the specific binding reaction between protein A and IgG (Fig. 3(a)).  (The details of the surface preparation are shown elsewhere. \cite{Zhou})  The measurement takes three steps.  First, we measured the through-thickness resonant vibration of the protein-A immobilized sensor film.  Second, we immersed the sensor film on an IgG solution with a concentration of 6.7 nM for 40 min and measured the reflectivity change after washing and drying the surfaces.  Third, we removed the IgG molecules from the surfaces by immersing the sensor chip in a glycine-HCl buffer (GHB) solution for 20 min and measured the reflectivity change after rinsing and drying the surfaces.  Figure 3(b) shows the reflectivity changes measured at the three steps, where the vibrational period appears to be longer by the addition of IgG molecules (see the inset), indicating that the resonant-frequency decreases because of the increase in the effective mass of the resonator.  Figure 3(c) shows the resonance frequencies of the 1st and 3rd modes.  They decreased by about 0.9\% when the target molecules (IgG) were captured on the surfaces and recovered by removing the IgG molecules using GHB.  Considering that the frequency change observed by an ultra-sensitive QCM biosensor for 5-nM IgG solution was 0.017\%, \cite{OgiAnalChem} the mass sensitivity of the free-standing diamond-film resonator developed here is more than 50 times higher than that of the most sensitive QCM biosensor.  

In conclusion, we performed the picosecond ultrasound measurement for free-standing diamond thin films and observed the significant reduction of the Brillouin-oscillation amplitude when the Pt thin film is deposited on both surfaces. This behavior was reproduced by the numerical calculation by considering the generation of the additional strain pulse at the bottom interface with a time difference and by introducing attenuation. It is also found that this phenomenon highly depends on the time delay and the diamond-film thickness. We eventually fabricated a diamond free-standing thin-film resonator that generates almost no Brillouin oscillations, and demonstrated the importance of this phenomenon by applying it to a mass-sensitive biosensor.

\section*{SUPPEMENTARY MATERIAL}
See supplementary material for an experiment for the graphite specimen (Fig. S1), strain-pulse forms for Fig. 2(b) (Fig. S2), additional numerical simulations (Fig. S3), and another experiment for a diamond specimen (Fig. S4).  Also, supplementary movies show details of the relationship between the reflectivity change and strain pulses.      

\section*{Acknowledgments}
This study was supported by JSPS KAKENHI Grant Nos. JP19H00862 and 20K21144. 

\section*{AUTHOR DECLARATIONS}
\subsection*{Conflict of Interest}
The authors have no conflicts to disclose.

\section*{DATA AVAILABILITY}
The data that support the findings of this study are available from the corresponding author upon reasonable request.

\end{document}